\documentstyle[10pt,emulateapj]{article}

\lefthead{Harris et al.}
\righthead{X-ray Emission from the Radio Jet in 3C 120}

\begin{document}

\title{X-ray Emission from the Radio Jet in 3C 120}

\author{D.E. Harris\altaffilmark{1}, J. Hjorth\altaffilmark{2},
A.C. Sadun\altaffilmark{3}, J.D. Silverman\altaffilmark{4}, \and
M. Vestergaard\altaffilmark{1,2}}






\altaffiltext{1}{CfA, 60 Garden St., Cambridge MA 02138;
harris@cfa.harvard.edu, mvestergaard@head-cfa.harvard.edu}

\altaffiltext{2}{Astronomical Observatory, University of Copenhagen, Juliane
Maries Vej 30, DK-2100 Copenhagen, Denmark; jens@astro.ku.dk}

\altaffiltext{3}{Department of Physics, University of Colorado at Denver, Campus
Box 157, P.O. Box 173364, Denver, CO  80217-3364; asadun@carbon.cudenver.edu}

\altaffiltext{4}{Dept. of Astronomy, University of Virginia, PO Box 3818,
Charlottesville, VA 22903-0818; jds6h@ariel.astro.virginia.edu}

\begin{abstract}
We report the discovery of X-ray emission from a radio knot at a
projected distance of 25$^{\prime\prime}$ from the nucleus of the
Seyfert galaxy, 3C 120.  The data were obtained with the ROSAT High
Resolution Imager (HRI).  Optical upper limits for the knot preclude a
simple power law extension of the radio spectrum and we calculate some
of the physical parameters for thermal bremsstrahlung and synchrotron
self-Compton models.  We conclude that no simple model is consistent
with the data but if the knot contains small regions with flat
spectra, these could produce the observed X-rays (via synchrotron
emission) without being detected at other wavebands.
\end{abstract}


\section{Introduction}

The emission process for the X-rays from knots and hotspots in
extragalactic radio jets is generally thought to arise from one (or
more) of three generally viable processes: thermal bremsstrahlung
(TB), synchrotron self-Compton (SSC), and synchrotron emissions.
Whilst quite convincing synchrotron and SSC models have been developed
for particular sources (e.g. M87 [Biretta, Stern, \& Harris 1991] and
Cygnus A [Harris, Carilli, \& Perley 1994]), TB models generally
require a gas density sufficient to produce observable Faraday effects
which are not seen.  For other sources (e.g. Pictor A and 3C 273
[R\"{o}ser, private communication]), none of these three processes is
satisfactory.

Since only a handful of radio knots and hotspots have been detected in
the X-ray band, more examples are continually sought to obtain
additional constraints on the physical parameters implied for each
type of model.  For this reason, we proposed ROSAT HRI observations of
3C 120 when optical emission was reported for a segment of the inner
radio jet (Hjorth et al. 1995).  Our purpose was to see if we could
detect the bright radio knot 4$^{\prime\prime}$ from the galactic
nucleus, plus the following segment of the radio jet which was the
section seen in the optical.  Although our data were beset with
problems which have made a detailed study of the area near the nucleus
difficult, we find X-ray emission coincident with the relatively faint
radio knot 25$^{\prime\prime}$ to the northwest of the nucleus (see
Fig 1).

In this paper we give the essential X-ray parameters from our
observation, briefly describe the processing techniques, discuss the
astrometry, and consider the emission process responsible for the
X-rays.  3C 120 is a Seyfert 1 radio galaxy at a redshift of 0.033.
We use a Hubble constant of 50 km~s$^{-1}$~Mpc$^{-1}$ which gives a distance of
200 Mpc and a scale of 0.91 kpc~arcsec$^{-1}$.

\section{Observations and Image Processing}

\subsection{The Data}

The observation consisted of two segments (hereafter ``A'' and ``B").
Segment A was observed between 1996 August 16 and September 12 in 24
observation intervals (``OBIs") for a total live time of 37,036 s.
Segment B was observed 1997 March 3-8 with 18 OBIs and 18,139 s.
Segment A is seriously compromised by problems with the aspect
determination of the satellite, resulting in an extended (east-west)
tear-drop shape for the strong X-ray emission from the nucleus of the
galaxy.  A temporal analysis demonstrated this to be caused by an
instrumental effect and since the nuclear X-ray emission is highly
variable (Halpern 1985) most of the emission must be unresolved; our
resolution is of order a few kpc whereas the apparent size of a source
varying on a year timescale must be less than one pc.  Segment B
(Figure 1) was much better, with only mild broadening of the HRI Point
Response Function (PRF).  Because of these differences we analyzed the
two segments separately.

\subsection{Ex post facto improvement of the effective PRF}

We have developed a method of improving HRI images which
contain a strong source and which suffer from aspect problems ascribed
to imperfections in the star tracker.  These techniques are described
in Harris et al. (1998a) and in that paper both segments of the 3C 120
observation were used as illustrative examples (see figs. 2 and 3 of
Harris et al. 1998a).  Since the primary aspect problem addressed is
believed to be associated with the ROSAT `wobbling' (a dither about
the nominal pointing position with a period of 402 s), the term
``dewobble" is used hereafter to describe the process of phase binning,
centroiding, shifting, and restacking.

\subsection{Astrometry for the X-ray emission from the 25$^{\prime\prime}$ radio knot}

The segment B X-ray map is superior to the longer exposure segment A
since the effective PRF (even after dewobbling) is smaller and more
circular.  Therefore, except for intensity measurements, we confine
our discussion to segment B.  Absolute positions from the ROSAT aspect
system are generally good to a few arcsec.  We find an offset between
the radio and X-ray cores of
$\Delta$~RA~=~0$\stackrel{\prime\prime}{\textstyle .}$75 and
$\Delta$~DEC~=~2$\stackrel{\prime\prime}{\textstyle .}$0
($\Delta$~r~=~2$\stackrel{\prime\prime}{\textstyle .}$1).  The
overlays of the X-ray and radio maps in this paper always shift the
X-ray map so as to align the nucleus with that of the radio.

The position of the X-ray knot relative to the nucleus is found to be
25$\stackrel{\prime\prime}{\textstyle .}$7 in position angle
-64.5$^{\circ}$.  To within the measuring errors (one arcsec and one
degree) these values are the same as those for the peak radio
brightness of the 25$^{\prime\prime}$ knot with respect to the radio
core although the actual X-ray peak is a bit closer to the outer edge
of the radio knot (see figure 2).  With this precision afforded by the
registration of the bright nucleus, and since there is no optical
candidate (m$_{V}~>$~19 mag.) for the X-ray emission, we are convinced
that the area containing the peak radio brightness and outer edge of
the 25$^{\prime\prime}$ knot is the correct identification for the
X-ray emission.

\placefigure{fig2}

\section{X-ray parameters of the 25$^{\prime\prime}$ knot}

As a check on our methods of intensity measurement we derived the
countrate and flux for the nuclear emission which is known to be
highly variable at X-ray and other wavelengths (Halpern 1985).  To
compare the current X-ray values with those of Halpern we measured the
counts with an aperture of r=19$^{\prime\prime}$ (at the radius of the
brightness minimum between the nucleus and the 25$^{\prime\prime}$
knot; see figure 1).  For the conversion from counts/s to flux, we
assumed a spectral model consistent with the findings of Halpern: the
galactic value of neutral hydrogen (log NH~=~21.035) and a power law
with energy index 0.7.  Our resulting flux values for the 2 to 10 keV
band (segment A: 6.66$\times$10$^{-11}$ ergs~s$^{-1}$~cm$^{-2}$;
segment B: 6.10$\times$10$^{-11}$ ergs~s$^{-1}$~cm$^{-2}$, with
statistical uncertainties of less than one percent for both values)
are slightly greater than the range of 2 to 5$\times$10$^{-11}$
erg~s$^{-1}$~cm$^{-2}$ found by Halpern in 1980 with the Einstein MPC.

The spatial geometry chosen to measure the intensity of the knot is
designed to minimize the effects of the wings of the PRF from the
strong nuclear emission and is based on a contour diagram of the
smoothed (FWHM=3$^{\prime\prime}$) segment B image.  Even though the
knot is not well defined in segment A, the same geometry was applied.
We defined an annulus centered on the nucleus with inner and outer
radii of 19$\stackrel{\prime\prime}{\textstyle .}$5 and
31$\stackrel{\prime\prime}{\textstyle .}$5 respectively.  The segment
of this annulus between PA~=~280$^{\circ}$ and 310$^{\circ}$ contains
the X-ray knot.  The inner radius of the annulus was chosen to lie at
the saddle point in X-ray brightness (between the core and the knot
distributions), and the overall size in both dimensions is close to
12$^{\prime\prime}$.  Thus we used the r=6$^{\prime\prime}$ scattering
correction (1.26; David et al. 1995).  The background brightness was
estimated from the remainder of the same annulus, except for segment A
where the pie section (30$^{\circ}$ wide) south of the knot was
rejected since it contained residual nuclear emission.

We measured both the original data and the dewobbled version.  For
segment B there was excellent agreement between the two versions,
whereas for segment A it was clear that the original map should not be
used because of the instrumental spreading of the nuclear emission.
The results are given in Table 1.  From the measured countrates, there
is no evidence for X-ray variability of the knot.  The FWHM of a
Gaussian fit to the 3$^{\prime\prime}$ smoothed map of the knot is
7$\stackrel{\prime\prime}{\textstyle .}$6 $\times$
6$\stackrel{\prime\prime}{\textstyle .}$0, slightly smaller than the
X-ray size of the nuclear emission, but entirely consistent with the
core size given the small number of counts defining the knot.  Thus
the X-ray knot is unresolved.

\placetable{tbl-1}

\section{Discussion}

\subsection{Spectrum}

Radio flux densities were measured from VLA maps kindly supplied by
R. C. Walker (the same data used in Walker et al. 1987).  All the
radio maps (1.4, 5, and 14.9 GHz) were constructed to have the same
beam size of 1$\stackrel{\prime\prime}{\textstyle .}$25.  We used two
rectangles to measure flux densities; one which included the whole
knot and the other which isolated the outer edge and peak of the knot.
The results are given in Table 2.

\placetable{tbl-2}

The optical (B and I bands) upper limits were obtained with a circular
aperture of diameter 4$^{\prime\prime}$ from the same data used in Hjorth et
al. (1995).  The measurements were corrected for galactic extinction
and are given in Table 3.

\placetable{tbl-3}
The spectrum of the knot is plotted in Figure 3.  From the optical
upper limits, it is clear that a single power law fit from the radio
data to the X-ray point is precluded.

\placefigure{fig3}

\subsection{A Thermal Model}

The difficulties generally encountered in applying a TB model to radio
hotspots are twofold.  The first problem is how to maintain an
over-pressured mass of hot gas outside a galaxy.  To explain the
observed X-ray emission with a TB model, a mass of some 10$^9$
M$_{\odot}$ at a temperature of order several 10$^6$\,K is required in a
volume characterized by a radius of a few kpc.  The resulting
pressures usually exceed those expected in the IGM by factors
significantly greater than one.  The second problem is that no
anomalous Faraday effects are observed in the radio polarization
studies, and with only small values of the presumed magnetic field,
both depolarization and anomalous Faraday rotation would be expected
from the purported hot gas (e.g. Cygnus A [Harris et al. 1994]; M87
[Biretta et al. 1991] ; 3C 390.3 [Harris, Leighly, \& Leahy 1998b]).

For the 3C 120 knot, a uniform sphere of gas with a radius of
3$^{\prime\prime}$ (2.7 kpc) would have the parameters listed in Table
4 in order to produce the observed X-ray intensity.  Since the
25$^{\prime\prime}$ knot is located at least 23 kpc from the center of
the galaxy, we think the thermal model entails improbable conditions
for the maintenance of the required over-dense gas as a discernible
entity.  Furthermore, Walker et al. (1987) state that ``Any Faraday
rotation or depolarization is too small to be detected in these
observations.''  A thermal model for the X-ray emission would most
likely produce observable polarization effects since the rotation
measure for a pathlength of 5.4 kpc with even a small magnetic field like
10~$\mu$G would be $\approx$9000~radians~m$^{-2}$ (Walker et al. find
upper limits of $\approx$50~rad~m$^{-2}$).  Thus, to sustain a TB
model, the gas would have to be behind the radio knot and aligned with
it by chance when viewed from the Earth.

\placetable{tbl-4}

\subsection{An SSC Model}

SSC emission is the process of choice for the radio hotspots of Cygnus
A (Harris et al. 1994).  In that case, the average magnetic field
strength required for the SSC model is entirely consistent with the
value found from the radio data alone under the classical assumption
of minimum energy conditions.  However, for the other known examples
of X-ray emission from radio hotspots, the SSC model fails because the
observed radio brightness is much too small to provide the required
photon densities.  This is true also for 3C 120.  We estimate the
photon energy density by extending the radio spectrum from 10$^{7}$ to
10$^{14}$~Hz with a power law of $\alpha$~=~0.65 (Walker et al. 1987).
For an emitting volume of a cylinder corresponding to the deconvolved
FWHM of the 6 cm radio map (r=0$\stackrel{\prime\prime}{\textstyle
.}$25, length=1$\stackrel{\prime\prime}{\textstyle .}$23) the
resulting photon energy density is 2$\times$10$^{-13}$~erg~cm$^{-3}$
and the ratio of inverse Compton to synchrotron emissions would be
0.004 (B $\approx$40~$\mu$G).  The predicted SSC flux density at 2 keV
is more than 4 orders of magnitude less than that observed.

\subsection{A Synchrotron Model}

Synchrotron models require relativistic electrons with Lorentz energy
factors of 10$^{7}$ in typical magnetic fields of order 100~$\mu$G.
For the case of knot A in the M87 jet, the synchrotron model for the
X-ray emission is preferred because the optical jet is almost
certainly caused by synchrotron emission, and the steepening of the
power law spectra in the optical, when extrapolated to the X-ray band,
are in reasonable accord with the observed X-ray intensity (Biretta et
al. 1991).  3C 390.3 (hotspot B) is another case for which the
synchrotron model appears to be favored (Harris et al. 1998b).

To get synchrotron X-ray emission from the brightest part of the radio
knot requires electrons with Lorentz energy factors of
4$\times$10$^{7}$ for a magnetic field strength of 40~$\mu$G (the
equipartition value).  These electrons would have a synchrotron
halflife of 132 years.  Our optical upper limits (figure 3) mean that
the power law connecting the radio and X-ray flux densities is not an
acceptable spectrum unless more than 2.4 mag of excess absorption is
present in the B band (more than 1.5 mag in the I band).  Note that
these excesses refer to a line of sight which passes no closer than 22
kpc to the nucleus and there is a high probability that the
25$^{\prime\prime}$ knot is closer to us than the galaxy because the
inner segment of the jet shows beaming effects.  For a
synchrotron model with the usual power law distribution of
relativistic electrons, it would be necessary to hypothesize the
presence of a flat spectrum component ($\alpha~\leq$~0.4) such that
the optical flux densities would lie below the observed upper limits
and the radio flux densities would be too weak to be isolated from the
surrounding emission.

Physically, such a feature might be a localized volume within the knot
where the shock structure is such as to extend the electron spectrum
to the higher energies required to produce X-ray synchrotron emission.
Kirk (1997) discusses how mildly relativistic shocks and oblique
magnetic fields can produce particle energy distributions with power
law spectra flatter than the canonical distribution from
non-relativistic shocks.

\placetable{tbl-5}

In Table 5 we give the synchrotron model requirements for four
possible volumes.  These parameters are reasonable for a small region
within the radio knot although the actual magnetic field values might
be governed by larger scale structures and thus differ from the
particular entries.  Although for most stationary non-thermal sources,
E$^2$ losses steepen the particle distribution at the higher energies,
the synchrotron halflives given in Table 5 are sufficiently long that
this effect is not required for transient phenomena.  The canonical
spectral steepening comes about for a constant injection spectrum
operating over times much longer than the loss time-scale at the
highest energies.  We envisage a process that arises from time to time
at various locations in the knot but that is not a single, long
lasting feature.

If these small regions are the correct explanation for the X-ray
emission from the 3C 120 knot, they might be present also in other
sources.  The observed decrease of X-ray intensity in knot A of the
M87 jet led Harris, Biretta, \& Junor (1997) to suggest that the X-ray
structure of knot A would be smaller than that in the radio and
optical.  The combined spectral and spatial resolutions of AXAF are
well suited to test our hypothesis.

\section{Summary}

With the possible exception of 3C 273, the one common characteristic
found for radio jet features which emit observable X-rays is a strong
gradient in the radio brightness.  For M87 knot A, this is the edge
facing up stream, toward the nucleus.  For 3C 390.3 it is the outer
edge of hotspot B where the jet appears to impinge on a dwarf galaxy
which is a member of the 3C 390.3 group.  In the case of 3C 120, the
observed X-ray location is also the site of a strong gradient in radio
brightness.  Given the normal interpretation of these features as
indications of a strong shock front, a synchrotron explanation of the
X-ray emission remains the most likely possibility even if it
necessitates the existence of one or more hitherto unnoticed
components.

\acknowledgments

We thank Craig Walker for supplying us with his radio maps of 3C 120
and H. Falcke, the referee, for useful comments which led to
improvements in the paper.  Work at the SAO was partially supported
by NASA contract NAS5-30934 and NASA grant NAG5-2960.



\figcaption{An overlay of the X-ray and radio maps.  The 5 GHz radio
map was provided by R. C. Walker from VLA observations (see Walker,
1997).  The X-ray map has been smoothed with a Gaussian of FWHM =
3$^{\prime\prime}$, and the peak intensity has been aligned with the
radio peak.  In units of 0.15 counts/pixel (pixels are
0$\stackrel{\prime\prime}{\textstyle .}$5), the contour levels are
drawn at 1, 2, 3, 4, 6, 8, 10, 15, 20, 25, 50, 75, 100, and
200. \label{fig1}}

\figcaption{The 25$^{\prime\prime}$ knot.  Only segment B of the
observations is used for the X-rays (greyscale).  The radio map was
provided by R. C. Walker from VLA observations (see Walker, 1997).  In
units of 30 $\mu$Jy/beam, the contour levels are $\pm$1, 2, 3, ... 10,
12, 14, ... 20.  The restoring beam size is
0$\stackrel{\prime\prime}{\textstyle .}$365. \label{fig2}}

\figcaption{The spectrum of the 25$^{\prime\prime}$ knot.  For the
radio data, we used the values measured with the smaller box centered
on the bright peak (Table 2).  The X-ray point is from Table 1 and the
optical upper limits from Table 3. \label{fig3}}

\vspace{-7cm}

\begin{deluxetable}{llcccc}
\tablecaption{X-ray intensity measurements of the 25$^{\prime\prime}$ knot \label{tbl-1}}

\tablehead{
\colhead{Observation}&\colhead{Net Counts}&\colhead{Countrate}&\colhead{Flux (0.2-2keV)}&\colhead{S(2keV)}&\colhead{Lx(0.2-2keV)} \\
&&\colhead{(counts~ks$^{-1}$)}&\colhead{(erg cm$^{-2}$ s$^{-1}$)}&\colhead{($\mu$Jy)}&\colhead{(ergs~s$^{-1}$)}
}
\startdata

segment A	&74.6$\pm$14	&2.5 (19\%)	&1.50$\times$10$^{-13}$&0.018
&7.16$\times$10$^{41}$ \nl
segment B	&35.5$\pm$7.2	&2.5 (20\%)	&1.45$\times$10$^{-13}$&0.017
&6.95$\times$10$^{41}$ \nl

\enddata

\tablecomments{The countrate column includes the appropriate
scattering correction of 1.26 and the indicated percentage errors (one
sigma) apply also to the following columns.  The flux column gives the
so called ``unabsorbed" flux which, for the chosen spectral model, is
1.8 times larger than the observed flux.  For the (unabsorbed) flux
density (S) a $\mu$Jy is equivalent to $10^{-29}$ ergs cm$^{-2}$
s$^{-1}$ Hz$^{-1}$.  The spectral model used to convert count rate to
flux and luminosity includes the galactic absorption (log NH = 21.035)
and a power law with energy index $\alpha$=0.7 (the power law which
connects the radio and the X-ray data).  For other spectral models,
the reader can consult the behavior of the relevant conversion factors
given in David et al. (1995).}
\end{deluxetable}
\begin{deluxetable}{cccc}
\tablecaption{Radio Flux Densities of the 25$^{\prime\prime}$ Knot \label{tbl-2}}
\tablehead{
\colhead{Frequency}&\colhead{S(total)}&\colhead{S(partial)}&\colhead{Peak(partial)}
\\
\colhead{(GHz)}&\colhead{(mJy)}&\colhead{(mJy)}&\colhead{(mJy/beam)}
}

\startdata
	1.452		&30.8&	17.4&	6.86 \nl
	4.985		&16.8&	 9.2&	3.41\nl
       14.915		&4.5&	 2.9&	1.42\nl

\enddata 

\tablecomments{The rectangle used for the total flux density of the
knot was 7$\stackrel{\prime\prime}{\textstyle .}$2$\times$
6$\stackrel{\prime\prime}{\textstyle .}$6.  For the peak brightness
and outer edge (designated ``partial"), the box was
3$\stackrel{\prime\prime}{\textstyle
.}$6$\times$4$\stackrel{\prime\prime}{\textstyle .}$5.}

\end{deluxetable}

\vspace{-1.0in}
\begin{table}
\dummytable\label{tbl-3}
\end{table}

\begin{table}

\begin{center}
\begin{tabular}{cc}
\multicolumn{2}{c}{Table 3. Upper Limits for the Optical Flux
Densities} \\ 
& \\
 \hline\hline
log Frequency&S \\
(Hz)&($\mu$Jy) \\ \hline

14.52&	$\leq$0.707 \\
14.834&	$\leq$0.181 \\ \hline
\end{tabular}
\end{center}

\end{table}

\begin{deluxetable}{ccccc}
\tablecaption{Thermal Bremsstrahlung Model Parameters \label{tbl-4}}

\tablehead{
	\colhead{kT}&\colhead{Unabsorbed Flux}&\colhead{Density}&\colhead{Pressure}&\colhead{Mass} \\	
	\colhead{(keV)}&\colhead{(erg~cm$^{-2}$~s$^{-1}$)}&\colhead{(cm$^{-3}$)}&\colhead{(dyne~cm$^{-2}$)}&\colhead{(M$_{\odot}$)}
}
\startdata
	1	&1.56$\times$10$^{-13}$	&0.23&7.1$\times$10$^{-10}$	&4.9$\times$10$^{8}$\nl
	3	&1.33$\times$10$^{-13}$	&0.21&1.9$\times$10$^{-9}$	&4.3$\times$10$^{8}$ \nl
	6	&1.29$\times$10$^{-13}$	&0.22&4.0$\times$10$^{-9}$	&4.6$\times$10$^{8}$ \nl
\enddata

\tablecomments{The calculations were performed on the basis of a uniform density gas
filling a sphere of radius 2.7 kpc (consistent with the HRI PRF for an
unresolved source).  For smaller radii the density and pressure
increase whereas the total mass decreases.  The flux listed is for the
0.2 to 2 keV band.}
\end{deluxetable}


\begin{deluxetable}{lrrcccr}

\tablecaption{Synchrotron Parameters for Flat Spectrum Components \label{tbl-5}}

\tablehead{
\colhead{Radius}&\colhead{Radius}&\colhead{B(min)}&\colhead{P(min)}&\colhead{log
E(tot)}&\colhead{$\gamma_{X}$}&\colhead{$\tau$}
\\
\colhead{(arcsec)}&\colhead{(pc)}&\colhead{($\mu$G)}&\colhead{(dynes~cm$^{-2}$)}&\colhead{(ergs)}&&\colhead{(yrs)}
}

\startdata
	0.5&450	 & 9&8$\times$10$^{-12}$&53.27&1.1$\times$10$^{8}$&750 \nl
	0.1& 91	 &38&1$\times$10$^{-10}$&52.37&5.5$\times$10$^{7}$&100 \nl
	0.05& 45& 70&4$\times$10$^{-10}$&52.05&4.1$\times$10$^{7}$& 42 \nl
	0.01& 9&277&7$\times$10$^{-9}$&51.15	&2.0$\times$10$^{7}$&  5 \nl

\enddata

\tablecomments{The basic spectrum is constructed from the observed X-ray flux density
with $\alpha$=0.4 and frequency limits of 100 MHz and 4.8$\times$10$^{17}$ Hz (2keV).
The total energy is that stored in relativistic particles and magnetic
fields.  $\gamma_{X}$ is the Lorentz energy factor of the electrons
responsible for the 2 keV emission and $\tau$ is the synchrotron
half-life of those electrons.}
\end{deluxetable}

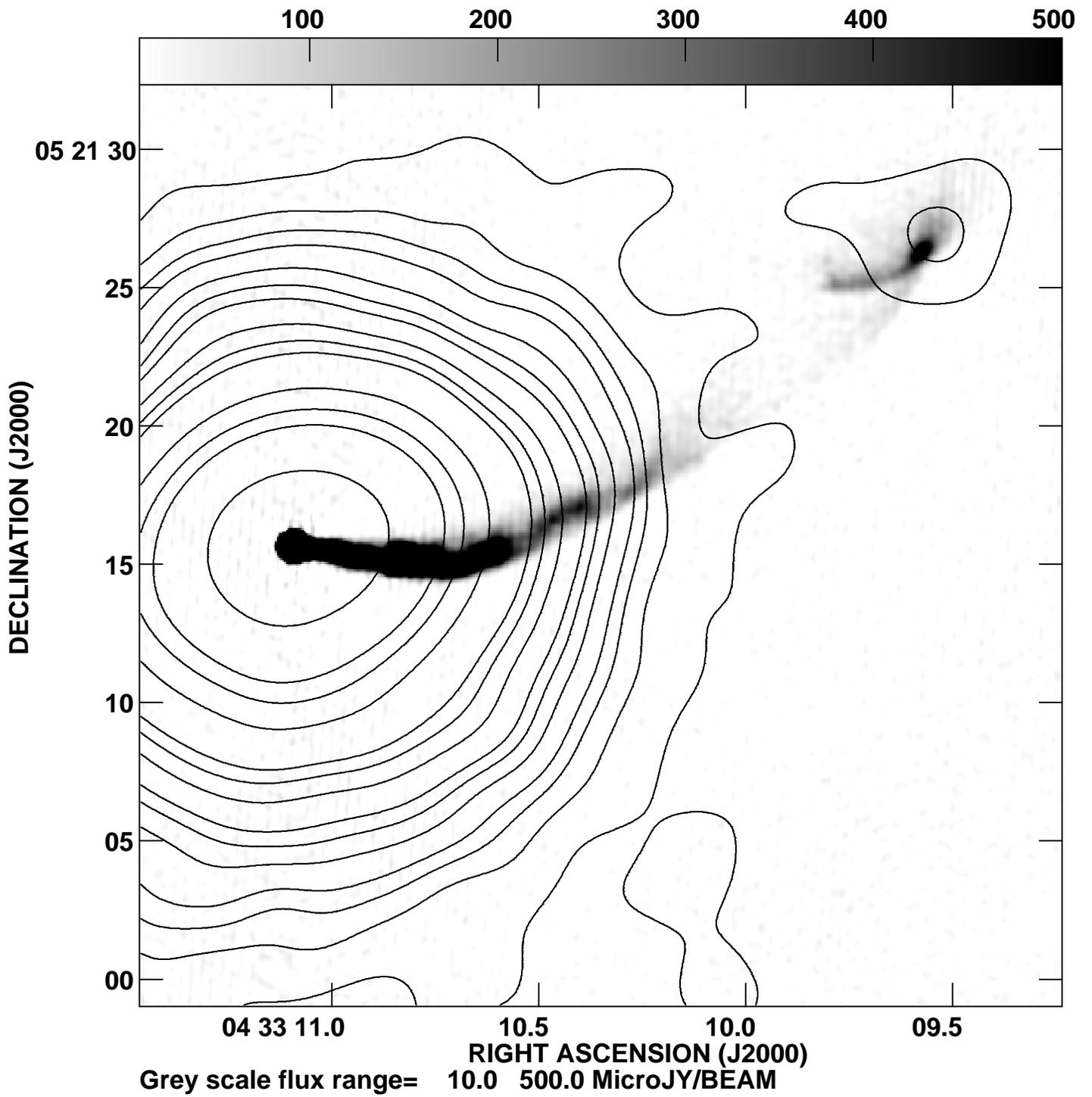
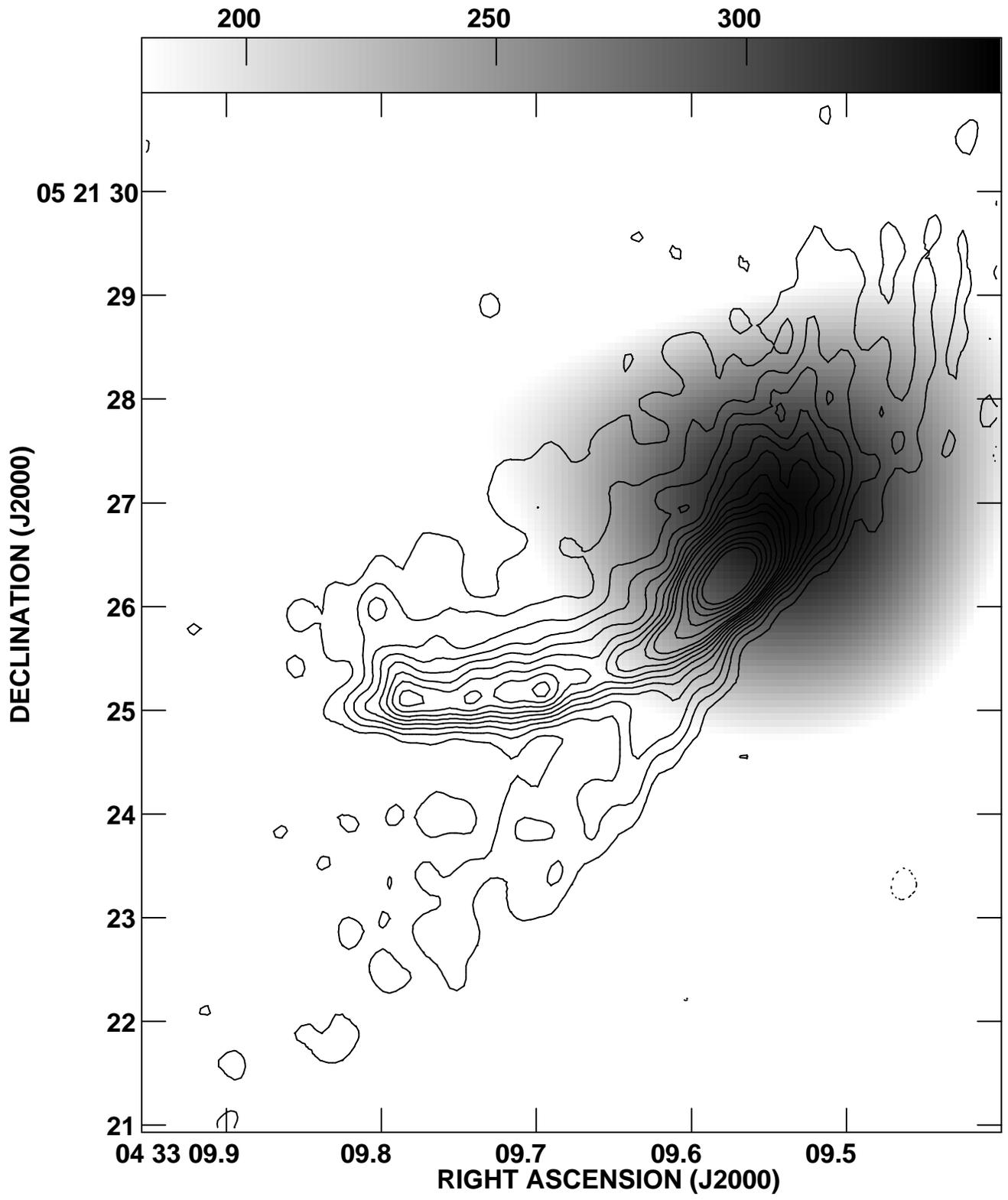
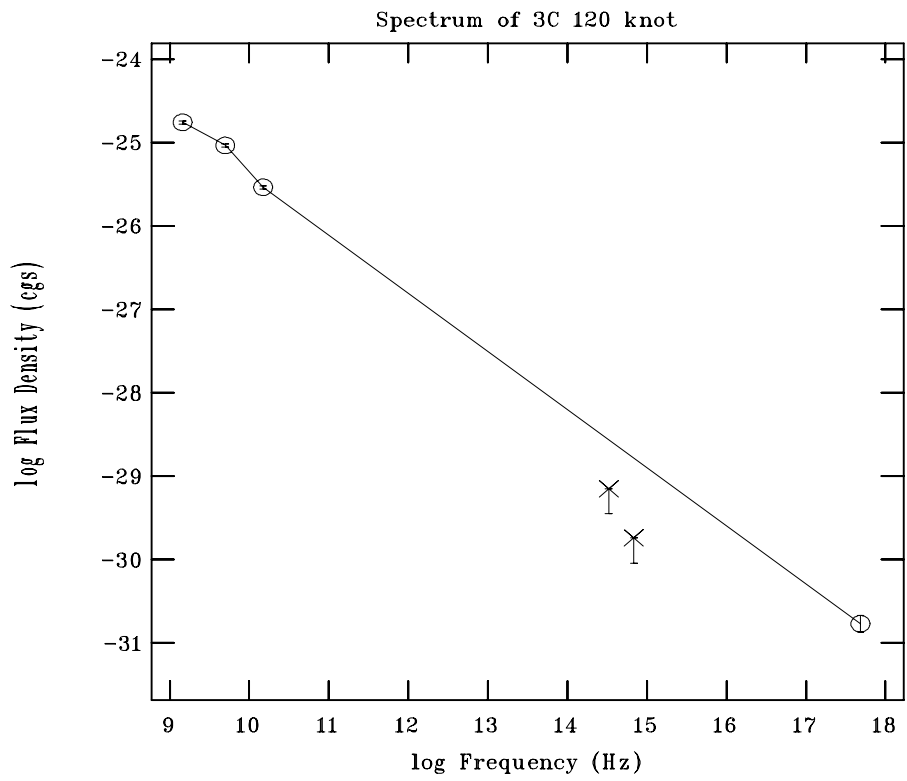
\end{document}